\documentclass[12pt,a4paper]{iopart}

\usepackage{graphicx}
\usepackage{datetime}
\usepackage{color}
\usepackage{mathrsfs} 
\usepackage{dsfont}
\usepackage{xspace}
\usepackage{iopams}  

\usepackage[bookmarks=true,
            bookmarksopen=true,colorlinks=false]{hyperref}

\newcommand{\eqref}{\eref}
\newcommand{\dd}{\mathrm{d}}
\newcommand{\ee}{\mathrm{e}}
\newcommand{\ii}{\mathrm{i}}

\newcommand{\E}{\mathcal E}

\newcommand{\Ap}{\mathcal A_+}
\newcommand{\Am}{\mathcal A_-}
\newcommand{\An}{\mathcal A_0}
\newcommand{\Ha}{{\mathcal H_1}}
\newcommand{\Hb}{{\mathcal H_2}}
\newcommand{\C}{{\mathcal C}}

\newcommand{\Y}{{\mathbf Y}}
\newcommand{\J}{{\mathbf J}}
\newcommand{\B}{{\mathbf B}}
\newcommand{\M}{{\mathbf M}}
\newcommand{\N}{{\mathbf N}}
\renewcommand{\P}{{\mathbf P}}
\newcommand{\Cmat}{{\mathbf C}}

\newcommand{\1}{{\mathbf 1}}
\newcommand{\0}{{\mathbf 0}}
\newcommand{\EE}{{\mathbf E}}
\renewcommand{\Re}{\mathrm{Re}}
\renewcommand{\Im}{\mathrm{Im}}

\makeatletter
\def\blfootnote{\xdef\@thefnmark{}\@footnotetext}
\makeatother

\begin{document}

\title
[On the balance problem for two rotating and charged black holes]
{On the balance problem for two rotating and charged black holes$^1$}

\author{J\"org Hennig}
\address{Department of Mathematics and Statistics,
           University of Otago,
           PO Box 56, Dunedin 9054, New Zealand}
\eads{\mailto{jhennig@maths.otago.ac.nz}}

\begin{abstract}
It is an interesting open problem whether two non-extremal rotating and electrically charged black holes can be in physical equilibrium, which might be possible due to a balance between the gravitational attraction and the spin--spin and electrical repulsions. Exact candidate solutions were constructed, but it is unclear whether they are physically acceptable. These solutions were obtained by assuming a particular behaviour on the symmetry axis. However, it was not known whether the assumed form of the axis data covers the general case or whether data of some other type need to be considered as well. By studying a boundary value problem for the axisymmetric and stationary Einstein--Maxwell equations, we address this question and derive the most general form of permissible axis potentials for possible equilibrium configurations.
\let\thefootnote\relax\footnotetext{This paper is dedicated to Gernot Neugebauer, to whom I  would like to offer my warmest thanks for introducing me to, and sharing his fascination for the intriguing world of general relativity in his lectures at the University of Jena many years ago, for having been a wonderful supervisor of my Diploma and PhD theses, and for the  delightful collaboration in our later joint projects.}
\let\thefootnote\svthefootnote
\\[2ex]{}
\end{abstract}


\section{Introduction\label{sec:intro}}

It is a classical result in Newtonian mechanics that (uncharged) $n$-body configurations cannot be in equilibrium for $n>1$ if the bodies are separated by a plane. Indeed, since the Newtonian gravitational force is always attractive, it is clear that separated bodies  cannot be in balance. Interestingly, the situation might be rather different in the nonlinear theory of general relativity. If we consider \emph{rotating} objects, then the effect of spin--spin repulsion might be able to compensate for the gravitational attraction. Therefore, in order to better understand the nature of the gravitational interaction in general relativity, it is an important question as to whether an equilibrium of (physically reasonable) rotating bodies is possible.

Probably the simplest type of such an equilibrium configuration might be a configuration consisting of two aligned rotating black holes in vacuum, i.e.\ a nonlinear superposition of two Kerr black holes. Especially since the discovery of a family of exact candidate solutions --- the \emph{double-Kerr-NUT solution} \cite{KramerNeugebauer1980,Neugebauer1980} --- such two-black-hole configurations have attracted great interest. It remained unclear, however, whether there is any choice of the parameters for which the double-Kerr-NUT solution does indeed describe regular spacetimes containing two black holes in equilibrium. Moreover, it was not a priori guaranteed that there could not be other candidate solutions that were not contained in this family. These questions were addressed in a series of papers \cite{NeugebauerHennig2009,HennigNeugebauer2011,NeugebauerHennig2012,Chrusciel2011} with the following results. Firstly, by studying a boundary value problem for an asymptotically flat spacetime with two black hole event horizons, it turned out that any regular solution would necessarily be a member of the double-Kerr-NUT family. Secondly, it was shown that none of the candidate solutions can describe a regular equilibrium configuration, for at least one of the two black holes would necessarily violate a geometric inequality between horizon area and angular momentum \cite{HennigAnsorgCederbaum2008}, which needs to be satisfied by physically reasonable black holes. Hence it turned out that stationary two-black-hole configurations in vacuum do not exist --- the spin--spin repulsion is not strong enough to compensate for the gravitational attraction.

What happens if we study the more general situation of \emph{electrovacuum} rather than vacuum solutions, i.e.\ if we allow for electromagnetic fields and consider possible equilibrium configurations with charged bodies? In non-relativistic physics, it is easily possible to construct charged $n$-body configurations. One simply needs to choose sufficiently large charges such that the electromagnetic forces cancel the gravitational forces. In the context of \emph{relativistic} two-black-hole configurations, however, there are upper limits for the charges (and rotation rates), since we otherwise obtain naked singularities instead of black holes. Hence it was not a priori guaranteed that there are any relativistic configurations where the combined spin--spin and charge--charge repulsions lead to stationary equilibrium.

Nevertheless, a class of static configurations is given by the well-known \emph{Majumdar--Papapetrou solution} \cite{Majumdar1947,Papapetrou1947}, which describes the superposition of an arbitrary number of extremal Reissner--Nordstr\"om black holes at arbitrary positions\footnote{A remarkable extension of the Majumdar--Papapetrou solution in the cosmological setting (with cosmological constant $\Lambda>0$) was constructed in \cite{KastorTraschen1993}. This solution contains an arbitrary number of black holes at fixed coordinate positions. However, due to the cosmological expansion, the proper distances between the black holes vary, i.e.\ it is not an equilibrium configuration in the context discussed here. 
Another interesting configuration is the superpartner of the Majumdar--Papapetrou solution in $N=2$ supergravity, which was derived in \cite{AichelburgEmbacher1986}. The black holes in this family of solutions carry angular momenta of a quantum mechanical nature, and the analysis of forces in \cite{KastorTraschen1999} shows that static configurations can exist due to a balance of the gauge and gravitational spin--spin forces.}.

Hence general relativity does evidently permit equilibrium states with charged black holes. 
However, this example requires \emph{extremal} black holes, i.e.\ degenerate horizons with vanishing surface gravity $\kappa$. According to the third law of black hole thermodynamics, however, it should not be possible by any procedure to reduce $\kappa$ to zero by any finite sequence of operations. In line with this principle is a result by Thorne \cite{Thorne1974}, who studied a black hole that swallows matter and radiation from an accretion disk. It turned out that the accreting matter can spin up the black hole up to a limiting state in which the ratio of the black hole rotation parameter and mass is about $0.998$ --- close to, but not quite at the extremal limit of $1$. Hence, while extremal black holes are certainly mathematically perfectly valid solutions to Einstein's field equations, they should rather be considered as limiting configurations that cannot exactly be realised in nature. Therefore, the above example for an equilibrium configuration of charged extremal black holes is most likely an unphysical idealisation, and the question remains as to whether equilibrium states with more realistic \emph{non-extremal} black holes are possible.

As far as \emph{static} solutions are concerned, this was answered in the negative. It was shown by Chru\'sciel and Tod \cite{ChruscielTod2007} that every static solution to the electrovacuum Einstein--Maxwell equations with disconnected horizons (i.e.\ multiple black holes) can only contain degenerate horizons. Moreover, such solutions are necessarily locally diffeomorphic to an open subset of a Majumdar--Papapetrou spacetime.

The problem is considerably more complicated in the case of \emph{rotating} black holes, i.e.\ non-static solutions, and it is currently not known whether physically reasonable equilibrium configurations do exist. Nevertheless, some families of exact candidate solutions were constructed \cite{ChamorroMankoSibgatullin1993,MankoMartinRuiz1994}. These were obtained by assuming that the solution (in terms of the Ernst potentials $\E$ and $\Phi$, see below) does have particular boundary values $\E_+(\zeta)$ and $\Phi_+(\zeta)$ on the upper part of the symmetry axis (above both black holes) in terms of a cylindrical coordinate $\zeta$. The chosen boundary values already determine the solution uniquely \cite{HauserErnst1981}, and the explicit solution in the entire spacetime was calculated by applying a particular technique from soliton theory (`Sibgatullin's integral method' \cite{Sibgatullin1984,MankoSibgatullin1993}).
A plausible form of the axis data $\E_+$ and $\Phi_+$ was obtained by starting from the Kerr--Newman data of a single black hole with mass $M$, rotation parameter $a$, and charge $Q$,
\begin{equation}
 \E_+(\zeta)=1-\frac{2M}{\zeta+M-\ii a}\equiv\frac{\zeta-M-\ii a}{\zeta+M-\ii a},\quad
 \Phi_+(\zeta)=\frac{Q}{\zeta+M-\ii a}
\end{equation}
and including additional terms to describe a second black hole. In \cite{ChamorroMankoSibgatullin1993}, boundary data of the form
\begin{equation}\fl\label{eq:bdata1}
 \E_+(\zeta)=1-\frac{2M_1}{\zeta+\zeta_1-\ii a_1}-\frac{2M_2}{\zeta+\zeta_2-\ii a_2},\quad
 \Phi_+(\zeta)=\frac{Q_1}{\zeta+\zeta_1-\ii a_1}+\frac{Q_2}{\zeta+\zeta_2-\ii a_2}
\end{equation}
were considered. On the other hand, the solution in \cite{MankoMartinRuiz1994} was constructed from the  data
\begin{eqnarray}
 \label{eq:bdata2a}
 \E_+(\zeta) &=&\frac{(\zeta+\zeta_1-M_1-\ii a_1)(\zeta+\zeta_2-M_2-\ii a_2)}
             {(\zeta+\zeta_1+M_1-\ii a_1)(\zeta+\zeta_2+M_2-\ii a_2)},\\
 \label{eq:bdata2b}
 \Phi_+(\zeta)&=&\frac{Q_1(\zeta+\zeta_2-\ii a_2)+Q_2(\zeta+\zeta_1-\ii a_1)}
               {(\zeta+\zeta_1+M_1-\ii a_1)(\zeta+\zeta_2+M_2-\ii a_2)}.
\end{eqnarray}
In both cases, the boundary values of the potentials are rational functions of $\zeta$ that depend on a number of free parameters.

In order to decide whether the candidate solutions  \cite{ChamorroMankoSibgatullin1993,MankoMartinRuiz1994} contain any physically reasonable equilibrium configurations, it needs to be studied whether the parameters can be chosen such that all of the following requirements are satisfied:
\begin{enumerate}
 \item the solutions have a vanishing NUT parameter (corresponding to the appropriate behaviour at infinity in an asymptotically flat spacetime),
 \item there are no conical singularities on the symmetry axes, in particular, between the two black holes (which would correspond to `struts' that keep the two black holes apart),
 \item the norm of the axial Killing vector vanishes on the symmetry axis, 
 \item there is no global magnetic charge,
 \item the solutions are free of singularities off the symmetry axis.
\end{enumerate}
Using the conditions (i)--(iv), one can write down an algebraic system of equations for the parameters that ensures the correct behaviour at infinity and on the axis. Unfortunately, the equations are rather involved, which makes it very difficult to decide whether there are parameter values satisfying those conditions. However, even if the correct behaviour on the axis is obtained in a subset of the parameter space, the solutions would likely violate condition (v), and it is probably even more difficult to check regularity off the axis. Hence it is not clear whether the solution families \cite{ChamorroMankoSibgatullin1993} and \cite{MankoMartinRuiz1994} contain any physically acceptable equilibrium configurations.

Moreover, the question remains as to whether the boundary data \eqref{eq:bdata1} or \eqref{eq:bdata2a} and \eqref{eq:bdata2b} do contain the axis potentials for actual equilibrium configurations with non-extremal rotating and charged black holes, if any exist, or whether data of some other form need to be considered. This is exactly the problem that we address in this paper. Generalising the considerations for one black hole in vacuum \cite{Neugebauer2000} (which leads to a constructive uniqueness proof of the Kerr solution), two black holes in vacuum \cite{NeugebauerMeinel,NeugebauerHennig2009,HennigNeugebauer2011,NeugebauerHennig2012}, or a single black hole in electrovacuum \cite{Meinel2012} (which extends the constructive uniqueness proof to the Kerr--Newman solution), we study a boundary value problem for two aligned charged and rotating black holes. As a result, we will obtain the most general form of the axis potentials $\E_+$ and $\Phi_+$.

These considerations crucially rely on the fact that the Einstein--Maxwell equations in electrovacuum for axisymmetric and stationary spacetimes can be reformulated in terms of a linear matrix problem, as a consequence of which methods from soliton theory are applicable. Note that closely related techniques also work in the context of Gowdy-symmetric cosmological models (which have two spacelike Killing vectors rather than a spacelike and a timelike Killing vector) \cite{HennigAnsorg2010,BeyerHennig2012,Hennig2016b,Hennig2019}. Yet another type of application is the investigation of the interior region of axisymmetric and stationary black holes \cite{AnsorgHennig2009,HennigAnsorg2009}.

This paper is organised as follows. In Sec.~\ref{sec:fieldeqns}, we recapitulate the Ernst formulation of the Einstein--Maxwell equations and the associated linear problem. Then we integrate the linear problem along the black hole horizons, symmetry axis and at infinity in Sec.~\ref{sec:integration}. This will eventually allow us to obtain the general form of the axis data. Finally, in Sec.~\ref{sec:discussion}, we summarise our results.

\section{Field equations\label{sec:fieldeqns}}
\subsection{Ernst formulation}

We describe the exterior electrovacuum region of an axisymmetric and stationary spacetime containing two aligned rotating and charged black holes with Weyl--Lewis--Papapetrou coordinates $(\rho,\zeta,\varphi,t)$. The line element can be written in the standard form
\begin{equation}\label{eq:metric}
 \dd s^2=f^{-1}
          \left[\ee^{2k}(\dd\rho^2+\dd\zeta^2)+\rho^2\,\dd\varphi^2\right]
         -f(\dd t+a\,\dd\varphi)^2,
\end{equation}
where the three metric functions $f$, $k$ and $a$ are functions of $\rho$ and $\zeta$ alone. The electromagnetic field can be given in terms of an electromagnetic 4-potential of the form $(A_\mu)=[0,0,A_\varphi(\rho,\zeta),A_t(\rho,\zeta)]$.

It is well-known that the corresponding Einstein--Maxwell equations can be written in a very elegant and concise form if we replace the metric functions and electromagnetic potential in terms of the two corresponding complex Ernst potentials $\E(\rho,\zeta)$ and $\Phi(\rho,\zeta)$ \cite{Ernst1968b}. The resulting \emph{Ernst equations} read
\begin{equation}\label{eq:Ernst}
 f\Delta\E   = (\nabla\E+2\bar\Phi\nabla\Phi)\cdot\nabla\E,\quad
 f\Delta\Phi = (\nabla\E+2\bar\Phi\nabla\Phi)\cdot\nabla\Phi,
\end{equation}
where $\Delta$ and $\nabla$ refer to the Laplace and nabla operators in flat cylindrical coordinates $(\rho,\zeta,\varphi)$, respectively, and a bar denotes complex conjugation. Note that the metric function $f$ is related to the Ernst potentials as follows,
\begin{equation}
 f=\Re(\E)+|\Phi|^2.
\end{equation}
Hence, if we define $b=\Im(\E)$, then the Ernst potential $\E$ can be expressed as
\begin{equation}\label{eq:b}
 \E=f-|\Phi|^2+\ii b.
\end{equation}
More details about the Ernst formulation of the field equations and the relation to the metric and electromagnetic functions can be found in \cite{Ernst1968b,Stephani}.

In our coordinates, the event horizon of a black hole is necessarily located at $\rho=0$ and corresponds to an interval on the $\zeta$-axis.
\begin{figure}\centering
 \includegraphics[width=4.5cm]{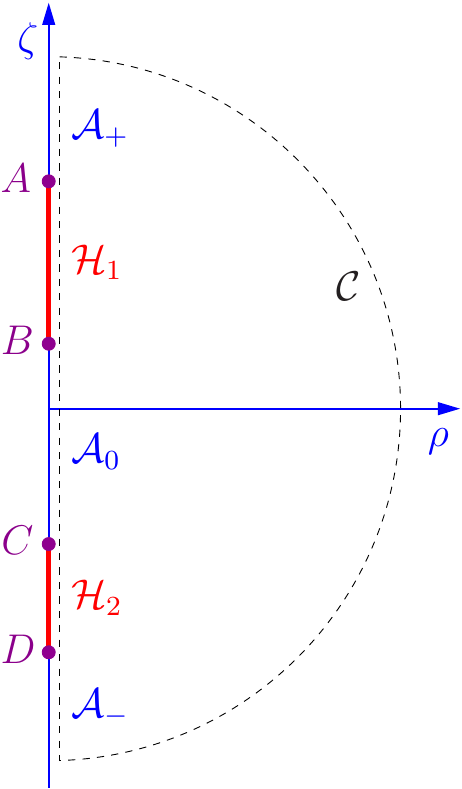}
 \caption{In Weyl--Lewis--Papapetrou coordinates, the event horizons $\Ha$ and $\Hb$ are located on the $\zeta$-axis. The symmetry axis has the three parts $\Ap$, $\An$ and $\Am$. In Sec.~\ref{sec:integration} below we will integrate the linear problem, which is equivalent to the field equations, along the dashed path. The part $\C$ of this path is an infinitely large semicircle.}
 \label{fig:IntegrationPath}
\end{figure}
The present situation of a (candidate) spacetime with two black holes is sketched in Fig.~\ref{fig:IntegrationPath}. We denote the endpoints of the horizons by $A$, $B$, and $C$, $D$, and the corresponding $\zeta$-intervals by $[K_B,K_A]$ and $[K_D,K_C]$, respectively.

\subsection{The linear problem}
It is a most remarkable property of the Ernst equations \eqref{eq:Ernst} that they belong to the class of integrable partial differential equations. They are equivalent to an associated \emph{linear} matrix problem, and techniques from soliton theory, like the inverse scattering method, can be used to study properties of the solutions and to construct exact solutions.

A linear problem (LP) for the electrovacuum Ernst equations was first found by Belinski \cite{Belinski1979}, and a modified version was constructed by Neugebauer and Kramer \cite{NeugebauerKramer}. Here we will use a minor reformulation of Neugebauer and Kramer's LP, which is due to Meinel \cite{Meinel2012}. In order to state the LP, we first define the complex coordinates
\begin{equation}
 z=\rho+\ii\zeta,\quad\textrm{and}\quad \bar z=\rho-\ii\zeta,
\end{equation}
and the function
\begin{equation}\label{eq:lambda}
 \lambda=\sqrt{\frac{K-\ii\bar z}{K+\ii z}},
\end{equation}
which depends on the complex coordinates and on an important additional degree of freedom, the \emph{spectral parameter} $K$. 
Due to the square root, the complex function $\lambda$ is defined on a two-sheeted Riemannian $K$-surface with branch points at $K_1=\ii\bar z$ and $K_2=-\ii z$.

The LP is a system of equations for a $3\times 3$ matrix function $\Y=\Y(\rho,\zeta;K)$ and reads
\begin{eqnarray}
\label{eq:LP1}
 \Y_{,z} &=& \left[
 \left(\begin{array}{ccc}
   B_1 & 0 & C_1\\ 0 & A_1 & 0\\ D_1 & 0 & 0
 \end{array}\right)
 +\lambda
 \left(\begin{array}{ccc}
  0 & B_1 & 0\\ A_1 & 0 & -C_1\\ 0 & D_1 & 0
 \end{array}\right)\right]\Y,\\
 \label{eq:LP2}
 \Y_{,\bar z} &=& \left[
 \left(\begin{array}{ccc}
   B_2 & 0 & C_2\\ 0 & A_2 & 0\\ D_2 & 0 & 0
 \end{array}\right)
 +\frac{1}{\lambda}
 \left(\begin{array}{ccc}
  0 & B_2 & 0\\ A_2 & 0 & -C_2\\ 0 & D_2 & 0
 \end{array}\right)\right]\Y.
\end{eqnarray}
The matrix elements are given in terms of the Ernst potentials by
\begin{eqnarray}
 A_1 &= \bar B_2 = \frac{1}{2f}(\E_{,z}+2\bar\Phi\Phi_{,z}),\quad
 C_1 &= f\bar D_2 = \Phi_{,z},\\
 A_2 &= \bar B_1 = \frac{1}{2f}(\E_{,\bar z}+2\bar\Phi\Phi_{,\bar z}),\quad
 C_2 &= f\bar D_1 = \Phi_{,\bar z}.
\end{eqnarray}
Note that integrability of the LP \eqref{eq:LP1}, \eqref{eq:LP2} is ensured by virtue of the Ernst equations \eqref{eq:Ernst}, since the integrability condition $\Y_{,z\bar z}=\Y_{,\bar z z}$ turns out to be equivalent to the Ernst equations.

Since the LP  contains $\lambda$, the matrix function $\Y$ will in general also take on different values on the two Riemannian $K$-sheets. Only at the branch points the function values are unique. If some function $\Y$ solves the LP on one sheet, then one can show that a particular solution on the other sheet is given by $\J\Y$ with
\begin{equation}
 \J=\mathrm{diag}(1,-1,1).
\end{equation}
The general solution on the other sheet can be obtained by multiplying $\J\Y$ on the right by a matrix that depends on $K$ only. However, only for a particular choice of this matrix, we obtain a solution that correctly connects to the solution on the first sheet through the branch cut. Hence the solutions on the two sheets are related via
\begin{equation}\label{eq:sheets}
 \Y|_{-\lambda}=\J\Y|_{\lambda}\B(K)
\end{equation}
for some $3\times 3$ matrix $\B$. It is possible to impose certain gauge conditions which  enforce that $\B$ takes on a particular form. For example, the matrix obtained with conditions used in \cite{Meinel2012} is given by $\B=\small\left(\begin{array}{ccc}0&1&0\\1&0&0\\0&0&1\end{array}\right)$. Here, however, we will demonstrate that the discussion can easily be done in full generality, i.e.\ without referring to any particular gauge. Indeed, the final physical results will, of course, be independent of any gauge choice. The only property of the matrix $\B$ that we will later use is that
\begin{equation}\label{eq:B2}
 \B^2=\1,
\end{equation}
where $\1$ is the $3\times 3$ identity matrix. This immediately follows by applying the transformation \eqref{eq:sheets} twice and using that this must lead back to the original solution $\Y$.

An important ingredient of our construction of two-black-hole solutions is to study the LP not only in the coordinates introduced above, but also in certain rotating frames of reference. They can be introduced by a simple transformation of the coordinate $\varphi$,
\begin{equation}
 \tilde\varphi=\varphi-\Omega t,
\end{equation}
where $\Omega$ is the angular velocity of the rotating frame. The other coordinates $\rho$, $\zeta$ and $t$ remain unchanged. In the following, we will consider the two particular frames that are co-rotating with either the first or the second black hole. If we denote the angular velocities of the two black holes by $\Omega_1$ and $\Omega_2$, then these frames correspond to choosing $\Omega=\Omega_{1/2}$. Note that we always assume rotating black holes with $\Omega_1\neq0$ and $\Omega_2\neq0$.

Fortunately, we do not need to solve the LP both in the original and the rotating frame, in order to obtain the two solutions $\Y$ and $\tilde\Y$. Instead, there is a simple relation between both solutions. The corresponding transformation in the vacuum case was given in \cite{NeugebauerMeinel}, and the generalisation to electrovacuum was presented in \cite{AnsorgHennig2009,HennigAnsorg2009}. In the present formulation of the LP, it reads \cite{Meinel2012}
\begin{equation}\fl\label{eq:rotframe}
 \tilde\Y(\rho,\zeta; K)=\left[
  \left(\begin{array}{ccc}
         c_- & 0   & 0\\
         0   & c_+ & 0\\
         0   & 0   & 1
        \end{array}\right)
  +\ii(K+\ii z)\frac{\Omega}{f}
  \left(\begin{array}{ccc}
         -1      & -\lambda & 0\\
         \lambda & 1        & 0\\
         0       & 0        & 0
        \end{array}\right)
\right]\Y(\rho,\zeta;K),
\end{equation}
where
\begin{equation}
 c_{\pm}=1+\Omega\left(a\pm\frac{\rho}{f}\right).
\end{equation}
The reason why using additional coordinate systems does actually add some new information is that the above transformation formula depends on the metric function $a$, which takes on specific boundary values at the symmetry axis and horizons (see next section). Considering not only $\Y$ but also $\tilde\Y$ does therefore incorporate these boundary conditions into our calculations.

\section{Integration of the linear problem\label{sec:integration}}

\subsection{Solutions on the axis parts and horizons}

Similarly to the study of two-black-hole configurations in vacuum and the other applications of the LP mentioned in the introduction, we intend to integrate the LP along the boundaries of the physical domain. The integration path consists of the event horizons $\Ha$ and $\Hb$ of the two black holes, the three parts $\Ap$, $\An$ and $\Am$ of the symmetry axis, and a semicircle $\C$ in the limit of an infinite radius, cf.\ Fig.~\ref{fig:IntegrationPath}.

In the following discussion, we will make use of the well-known boundary values for the metric and Ernst potentials at symmetry axes and black hole (Killing) horizons in Weyl--Lewis--Papapetrou coordinates, as well as the behaviour at infinity,
\begin{eqnarray}
 \Ap,\ \An,\ \Am:\quad && a=0,\label{eq:condA}\\
 \Ha:                  && a=-\frac{1}{\Omega_1},\label{eq:condH1}\\
 \Hb:                  && a=-\frac{1}{\Omega_2},\label{eq:condH2}\\
 A,\ B,\ C,\ D:        && f=0,\label{eq:condf}\\
 \C:                   && \E\to 1,\quad \Phi\to 0.\label{eq:condC}
\end{eqnarray}
Here, $A$, $B$, $C$, $D$ refers to the endpoints of the horizons, see Fig.~\ref{fig:IntegrationPath}.

Firstly, we consider the LP anywhere on the $\zeta$-axis, i.e.\ at $\rho=0$. According to \eqref{eq:lambda}, the function $\lambda$ simplifies to $\lambda=\pm1$ on the two Riemannian sheets. As a consequence, the LP \eqref{eq:LP1}, \eqref{eq:LP2} also becomes particularly simple and reduces to an ODE. The general solution can easily be derived. In the sheet with $\lambda=1$ it reads
\begin{equation}
 \Y(0,\zeta; K)=\EE(\zeta)\Cmat(K),\quad
 \EE:=\left(\begin{array}{ccc}
             \bar\E+2|\Phi|^2 & 1 & \Phi\\
             \E & -1 & -\Phi\\
             2\bar\Phi & 0 & 1
            \end{array}\right).
\end{equation}
Hence $\Y$ depends on the boundary values of the Ernst potentials and on a $K$-dependent `integration constant', a $3\times 3$ matrix $\Cmat$. The solution in the other sheet with $\lambda=-1$ is readily obtained from \eqref{eq:sheets}.

Using \eqref{eq:rotframe}, we can also construct the solution in the frame that rotates with angular velocity $\Omega$. The result is
\begin{equation}\fl
 \tilde\Y = \left[\left(\begin{array}{ccc}
                    1+\Omega a & 0 & 0\\
                    0 & 1+\Omega a & 0\\
                    0 & 0 & 1
                   \end{array}\right)\EE
        +2\ii\Omega(K-\zeta)\left(\begin{array}{ccc}
                                   -1 & 0 & 0\\
                                   1 & 0 & 0\\
                                   0 & 0 & 0
                                  \end{array}\right)\right]\Cmat.
\end{equation}
This expression simplifies further if we specialise to the symmetry axis or horizons, using the boundary conditions \eqref{eq:condA}, \eqref{eq:condH1}, \eqref{eq:condH2}, where we consider the co-rotating frames with $\Omega=\Omega_1$ or $\Omega=\Omega_2$.

Now we can write down the expressions for $\Y$ and $\tilde\Y$ on the three parts of the symmetry axis and on the two horizons. In terms of $K$-dependent $3\times 3$ matrices $\Cmat_+$, $\Cmat_-$, $\Cmat_0$, $\Cmat_1$ and $\Cmat_2$, we have for $\lambda=+1$,
\begin{eqnarray}
 \label{eq:ApY}
 \Ap:\quad && \Y=\EE\Cmat_+,\\ 
 \label{eq:ApYs}
           && \tilde\Y=\left[\EE+2\ii\Omega_{1}(K-\zeta)\left(\begin{array}{ccc}
                                                      -1 & 0 & 0\\
                                                      1 & 0 & 0\\
                                                      0 & 0 & 0
                                                     \end{array}\right)\right]\Cmat_+,\\[1.5ex]
 \An:\quad && \Y=\EE\Cmat_0,\\
           && \tilde\Y=\left[\EE+2\ii\Omega_{1/2}(K-\zeta)\left(\begin{array}{ccc}
                                                      -1 & 0 & 0\\
                                                      1 & 0 & 0\\
                                                      0 & 0 & 0
                                                     \end{array}\right)\right]\Cmat_0,\\[1.5ex]                                                  
 \label{eq:AmY}
 \Am:\quad && \Y=\EE\Cmat_-,\\
           && \tilde\Y=\left[\EE+2\ii\Omega_{2}(K-\zeta)\left(\begin{array}{ccc}
                                                      -1 & 0 & 0\\
                                                      1 & 0 & 0\\
                                                      0 & 0 & 0
                                                     \end{array}\right)\right]\Cmat_-,\\[1.5ex]   
 \label{eq:H1Y}
 \Ha:\quad && \Y=\EE\Cmat_1,\\
 \label{eq:H1Ys}
           && \tilde\Y=\left[\left(\begin{array}{ccc}
                               0 & 0 & 0\\ 0 & 0 & 0\\ 0 & 0 & 1
                              \end{array}\right)
\EE+2\ii\Omega_{1}(K-\zeta)\left(\begin{array}{ccc}
                                                      -1 & 0 & 0\\
                                                      1 & 0 & 0\\
                                                      0 & 0 & 0
                                                     \end{array}\right)\right]\Cmat_1,\\[1.5ex] 
\Hb:\quad && \Y=\EE\Cmat_2,\\
           && \tilde\Y=\left[\left(\begin{array}{ccc}
                               0 & 0 & 0\\ 0 & 0 & 0\\ 0 & 0 & 1
                              \end{array}\right)
\EE+2\ii\Omega_{2}(K-\zeta)\left(\begin{array}{ccc}
                                                      -1 & 0 & 0\\
                                                      1 & 0 & 0\\
                                                      0 & 0 & 0
                                                     \end{array}\right)\right]\Cmat_2. 
\end{eqnarray}
Note that we consider \emph{both} co-rotating frames with $\Omega=\Omega_1$ and $\Omega=\Omega_2$ on the axis part $\An$, but otherwise only that co-rotating frame with the angular velocity of the nearest horizon. Again, the expressions in the Riemannian sheet with $\lambda=-1$ can be obtained from the above equations using \eqref{eq:sheets}.

Secondly, we consider the LP on the infinitely large semicircle $\C$. A semicircle with finite radius $R$ can be parametrised by $\rho=R\sin\alpha$, $\zeta=R\cos\alpha$, $0\le\alpha\le\pi$. On this semicircle, the function $\lambda$ becomes
\begin{equation}
 \lambda=\sqrt{\frac{K-\ii R(\sin\alpha-\ii\cos\alpha)}{K+\ii R(\sin\alpha+\ii\cos\alpha)}},
\end{equation}
which simplifies to $\lambda=\pm\ee^{\ii\alpha}$ in the limit $R\to\infty$. Hence, if we start on $\Ap$ with $\lambda=+1$, then the semicircle $\C$ leads us to the sheet on $\Am$ with $\lambda=-1$, and vice versa. This will be important later, when we continuously connect the solutions on the various parts of the boundary. Then we need to compare the $\lambda=1$ solution on $\Ap$ with the $\lambda=-1$ solution on $\Am$.

If we consider the LP on $\mathcal C$, using the asymptotic behaviour \eqref{eq:condC} of the Ernst potentials, we simply obtain $\Y_{,z}=\0$ and $\Y_{,\bar z}=\0$. Therefore, $\Y$ is constant on $\C$. (More precisely, the Ernst potentials in an asymptotically flat spacetime approach their constant limits at infinity at a rate for which the coefficient matrices on the right-hand side of the LP are of $\mathcal O (R^{-2})$ on a semicircle with coordinate radius $R$, while the length of the semicircle only increases proportional to $R$ as $R\to\infty$.)

Next we have a closer look at the various $\Cmat$-matrices that appear in the solution to the LP on the different parts of the boundary. These matrices cannot be chosen independently of each other. Instead we have to ensure that the solutions $\Y$ and $\tilde\Y$ are continuous at the points $A$, $B$, $C$, $D$, cf.\ Fig.~\ref{fig:IntegrationPath}, and that the solutions on $\Ap$ and $\Am$ are correctly connected via $\C$ as discussed above.

We start by considering continuity of $\Y$ at point $A$. For $\lambda=1$, using \eqref{eq:ApY} and \eqref{eq:H1Y}, we obtain the condition
\begin{equation}\label{eq:Acont}
 \EE\Cmat_+=\EE\Cmat_1\quad\textrm{at}\quad \rho=0,\ \zeta=K_A.
\end{equation}
The same condition also ensures continuity of $\Y$ in the sheet $\lambda=-1$. Note that the nine components of the matrix condition \eqref{eq:Acont} are not independent. Instead, the second row is the negative of the first row. Hence we will only use the second and third row conditions.

Similarly, considering continuity of $\tilde\Y$ in the frame with $\Omega=\Omega_1$, we obtain the condition (for $\lambda=1$)
\begin{eqnarray}\
 \fl
 \left[\EE+2\ii\Omega_{1}(K-\zeta)\left(\begin{array}{ccc}
                                                      -1 & 0 & 0\\
                                                      1 & 0 & 0\\
                                                      0 & 0 & 0
                                                     \end{array}\right)\right]\Cmat_+
                                                     \nonumber\\
 \fl
 \quad=\left[\left(\begin{array}{ccc}
                               0 & 0 & 0\\ 0 & 0 & 0\\ 0 & 0 & 1
                              \end{array}\right)
\EE+2\ii\Omega_{1}(K-\zeta)\left(\begin{array}{ccc}
                                                      -1 & 0 & 0\\
                                                      1 & 0 & 0\\
                                                      0 & 0 & 0
                                                     \end{array}\right)\right]\Cmat_1
                             \quad\textrm{at}\quad \rho=0,\ \zeta=K_A.
\end{eqnarray}
Again, the nine components are not independent: the second row is the negative of the first row. Moreover, the third row is identical with the third row of \eqref{eq:Acont}. Hence we obtain one new row of conditions.

The three rows of independent conditions can be combined as follows,
\begin{equation}\fl
 \left(\begin{array}{ccc}
        \E_A                     & -1 & -\Phi_A\\
        \E_A+2\ii\Omega_1(K-K_A) & -1 & -\Phi_A\\
        2\bar\Phi_A              &  0 & 1
       \end{array}\right)\Cmat_+
 =\left(\begin{array}{ccc}
        \E_A                     & -1 & -\Phi_A\\
        2\ii\Omega_1(K-K_A)      &  0 & 0\\
        2\bar\Phi_A              &  0 & 1
       \end{array}\right)\Cmat_1,
\end{equation}
where, here and in the following, subscripts $A$, $B$, $C$, $D$ refer to evaluation of function values at the indicated points, i.e.\ at $\rho=0$ and $\zeta=K_A,\ K_B,\ K_C$, or $K_D$, respectively. Solving for $\Cmat_1$, we can also write this condition in the form
\begin{equation}\label{eq:condC1}
 \Cmat_1=\left(\1+\frac{1}{\alpha_A}\M_A\right)\Cmat_+,
\end{equation}
where
\begin{equation}
 \alpha_A:=2\ii\Omega_1(K-K_A)
\end{equation}
and
\begin{equation}\label{eq:MA}
 \M_A:=m_A n_A^T,\quad
 m_A:=\left(\begin{array}{c}
             -1\\ \bar\E_A\\ 2\bar\Phi_A
            \end{array}\right),\quad
 n_A:=\left(\begin{array}{c}
             -\E_A\\ 1\\ \Phi_A
            \end{array}\right).
\end{equation}
Note that $n_A\cdot m_A\equiv n_A^T m_A=0$ [cf.~\eqref{eq:condf}], which implies $\M_A^2=\0$. As a consequence, matrices of the form $\1+c\M_A$ can easily be inverted, and we have
$(\1+c\M_A)^{-1}=\1-c\M_A$.

Repeating the above calculations at the points $B$, $C$, and $D$, we obtain the additional conditions
\begin{eqnarray}
  \label{eq:condC0}
  \Cmat_0 &=& \left(\1-\frac{1}{\alpha_B}\M_B\right)\Cmat_1,
   \quad \alpha_B:=2\ii\Omega_1(K-K_B),\\
  \label{eq:condC2}
  \Cmat_2 &=& \left(\1+\frac{1}{\alpha_C}\M_C\right)\Cmat_0,
   \quad \alpha_C:=2\ii\Omega_2(K-K_C),\\
  \label{eq:condCm}
  \Cmat_- &=& \left(\1-\frac{1}{\alpha_D}\M_D\right)\Cmat_2,
   \quad \alpha_D:=2\ii\Omega_2(K-K_D),
\end{eqnarray}
where the matrices $\M_B$, $\M_C$, $\M_D$ are defined as in \eqref{eq:MA}, but with the Ernst potentials evaluated at the points $B$, $C$, or $D$, respectively.

If we would know the exact form of the matrix $\Cmat_+$, then the above conditions would allow us to compute all of the remaining $\Cmat$-matrices. This would indeed be possible if we imposed suitable gauge conditions for the LP, see \cite{Meinel2012}. Here, however, as mentioned before, we intend to demonstrate that the final results can easily be obtained without any particular gauge.

Finally, we consider the transition from the solution on $\Ap$ to $\Am$ via $\C$. Based on our earlier discussion, we arrive at the condition 
\begin{equation}
 \lim_{\zeta\to\infty}\Y(0,\zeta;K)|_{\lambda=1}
 =\lim_{\zeta\to-\infty}\Y(0,\zeta;K)|_{\lambda=-1}.
\end{equation}
With the explicit solutions \eqref{eq:ApY} and \eqref{eq:AmY}, together with \eqref{eq:sheets} and the asymptotic values \eqref{eq:condC}, the previous equation becomes
\begin{equation}
 \left(\begin{array}{ccc}
        1 & 1 & 0\\
        1 &-1 & 0\\
        0 & 0 & 1
       \end{array}\right)\Cmat_+
 =\J\left(\begin{array}{ccc}
        1 & 1 & 0\\
        1 &-1 & 0\\
        0 & 0 & 1
       \end{array}\right)\Cmat_-\B.
\end{equation}
This can be rearranged to
\begin{equation}\label{eq:cond2}
 \Cmat_-\B=\P\Cmat_+,
\end{equation}
where $\P$ is the following permutation matrix,
\begin{equation}
 \P:=\left(\begin{array}{ccc}
            0 & 1 & 0\\
            1 & 0 & 0\\
            0 & 0 & 1
           \end{array}\right).
\end{equation}

\subsection{Parameter conditions}

The solutions of the LP at the symmetry axis and horizon discussed in the previous subsection depend on the values of the Ernst potentials at the points $A$, $B$, $C$, $D$, the $\zeta$-coordinates $K_A$, $K_B$, $K_C$, $K_D$ of these points, and the angular velocities $\Omega_1$, $\Omega_2$ of the two horizons. These parameters, however, cannot be chosen independently of each other. Instead, we have to impose a number of parameter conditions. In the following, we show how these conditions can be obtained.

Combining \eqref{eq:condC1}, \eqref{eq:condC0}, \eqref{eq:condC2}, \eqref{eq:condCm}, we obtain an equation relating $\Cmat_-$ and $\Cmat_+$,
\begin{equation}\fl
 \Cmat_-=\left(\1-\frac{1}{\alpha_D}\M_D\right)
         \left(\1+\frac{1}{\alpha_C}\M_C\right)
         \left(\1-\frac{1}{\alpha_B}\M_B\right)
         \left(\1+\frac{1}{\alpha_A}\M_A\right)\Cmat_+.
\end{equation}
Another relation between $\Cmat_-$ and $\Cmat_+$ was previously obtained in \eqref{eq:cond2}. Combining those equations, we can eliminate $\Cmat_-$ and solve for $\B$,
\begin{equation}\fl\label{eq:Cpluscond}
 \B=\Cmat_+^{-1}
     \left(\1-\frac{1}{\alpha_A}\M_A\right)
     \left(\1+\frac{1}{\alpha_B}\M_B\right)
     \left(\1-\frac{1}{\alpha_C}\M_C\right)
     \left(\1+\frac{1}{\alpha_D}\M_D\right)\P\Cmat_+.
\end{equation}
Plugging this into the relation $\B^2=\1$ [cf.~\eqref{eq:B2}], we obtain 
\begin{eqnarray}
 \fl
  \P
  \left(\1-\frac{1}{\alpha_D}\M_D\right)
  \left(\1+\frac{1}{\alpha_C}\M_C\right)
  \left(\1-\frac{1}{\alpha_B}\M_B\right)
  \left(\1+\frac{1}{\alpha_A}\M_A\right)\nonumber\\
  \fl\quad
  =
  \left(\1-\frac{1}{\alpha_A}\M_A\right)
  \left(\1+\frac{1}{\alpha_B}\M_B\right)
  \left(\1-\frac{1}{\alpha_C}\M_C\right)
  \left(\1+\frac{1}{\alpha_D}\M_D\right)\P.
\end{eqnarray}
As expected, the gauge dependent matrices $\B$ and $\Cmat_+$ have cancelled, and hence the physical restrictions are independent of any gauge choice.
Finally, simplifying and multiplying both sides by $\alpha_A\alpha_B\alpha_C\alpha_D$, we obtain the condition that two matrix polynomials of third degree in $K$ must be the same. Equating the coefficients of $K^0$, $K^1$, $K^2$ and $K^3$ leads to a number of constraints for the parameters.

Note that the corresponding conditions in the case of a single black hole in electrovacuum can easily be solved explicitly \cite{Meinel2012}. In the present case, unfortunately, they are much more involved, and an explicit solution may be difficult to obtain. Fortunately, however, most of the conditions and their solution are not required in the following.

As an example, we only give the three simplest conditions here, which are
\begin{equation}\label{eq:parcond}
 \Omega_1(|\E_C|^2-|\E_D|^2)+\Omega_2(|\E_A|^2-|\E_B|^2)=0,
\end{equation}
\begin{equation}
 \Omega_1(\E_C+\bar\E_C-\E_D-\bar\E_D)+\Omega_2(\E_A+\bar\E_A-\E_B-\bar\E_B)=0,
\end{equation}
\begin{equation}
 \Omega_1[(1-\bar\E_C)\Phi_C-(1-\bar\E_D)\Phi_D]
 +\Omega_2[(1-\bar\E_A)\Phi_A-(1-\bar\E_B)\Phi_B]=0.
\end{equation}
For our derivation of the most general form of the axis potentials on $\Ap$ in the next subsection, we will explicitly only require Eq.~\eqref{eq:parcond}.

\subsection{Construction of the axis potentials}

With the preparations from the previous subsection, we are now in a position to obtain the Ernst potentials $\E_+=\E(0,\zeta)$ and $\Phi_+=\Phi(0,\zeta)$ on $\Ap$. The key ingredient for this construction is the above-mentioned property that $\Y$ can generally take on different values on the two Riemannian sheets, but must be unique at the branch points $K_1=\ii\bar z=\zeta+\ii\rho$ and $K_2=-\ii z=\zeta-\ii\rho$ where the two sheets are connected. In the limit $\rho\to0$, as we approach the $\zeta$-axis, both branch points converge to $K_1=K_2=\zeta$, i.e.\ we have confluent branch points.

With \eqref{eq:ApY} and \eqref{eq:sheets}, the condition that $\Y$ for $\lambda=1$ and $\Y$ for $\lambda=-1$ on $\Ap$ coincide at $K=\zeta$ becomes
\begin{equation}
 \EE\Cmat_+=\J\EE\Cmat_+\B  \quad\textrm{at}\quad K=\zeta.
\end{equation}
Using \eqref{eq:B2}, we can rewrite this equation as
\begin{equation}\label{eq:PCBC}
 \Cmat_+\B\Cmat_+^{-1}\P=\EE^{-1}\J\EE\P     \quad\textrm{at}\quad K=\zeta.
\end{equation}
Now we define\footnote{Note that $\N$ generalises the $2\times2$ matrix $\mathcal N$ used in the discussion of two-black-hole configurations in vacuum, see Eq.~(22) in \cite{NeugebauerHennig2009}.}
\begin{equation}\fl
 \N(\zeta):=\EE^{-1}\J\EE\P
 \equiv
 \frac{1}{f_+}\left(\begin{array}{ccc}
                    1                  & |\Phi_+|^2-\ii b_+  & \Phi_+\\
                    |\Phi_+|^2+\ii b_+ & |\E_+|^2            & -\Phi_+\bar\E_+\\
                    -2\bar\Phi_+       & 2\bar\Phi_+\E_+     & f_+-2|\Phi_+|^2
                   \end{array}\right),
\end{equation}
where $b=\Im(\E)$, cf.~\eqref{eq:b}. If we reformulate the parameter condition \eqref{eq:Cpluscond} in the form
\begin{equation}\fl
 \Cmat_+\B\Cmat_+^{-1}\P=
     \left(\1-\frac{1}{\alpha_A}\M_A\right)
     \left(\1+\frac{1}{\alpha_B}\M_B\right)
     \left(\1-\frac{1}{\alpha_C}\M_C\right)
     \left(\1+\frac{1}{\alpha_D}\M_D\right)
\end{equation}
and specialise to $K=\zeta$, then, using \eqref{eq:PCBC}, we see that the matrix $\N$, which contains various combinations of the Ernst potentials, can be obtained from
\begin{equation}\fl\label{eq:N}
 \N= \left.\left(\1-\frac{1}{\alpha_A}\M_A\right)
     \left(\1+\frac{1}{\alpha_B}\M_B\right)
     \left(\1-\frac{1}{\alpha_C}\M_C\right)
     \left(\1+\frac{1}{\alpha_D}\M_D\right)\right|_{K=\zeta}.
\end{equation}
Note that the components of the matrix on the right-hand side simplify to rational functions in $\zeta$ with polynomials of at most fourth degree. 
Moreover, similarly to the parameter conditions, the gauge dependent matrices $\B$ and $\Cmat_+$ do not appear in this formula, so the axis potentials are certainly independent of any gauge choice for the LP.

Now we consider the following combinations of components of $\N$, where we evaluate the right-hand side of \eqref{eq:N} in each case, in order to determine the polynomial structure of the relevant components,
\begin{eqnarray}
 \fl\label{eq:f+}
 f_+ &=& \frac{1}{N_{11}} = \frac{\pi_4(\zeta)}{p_4(\zeta)},
 \hspace{2cm}  \pi_4:=(\zeta-K_A)(\zeta-K_B)(\zeta-K_C)(\zeta-K_D)\\
 \fl
     &&\hspace{4.52cm}p_4:\ \mbox{real monic polynomial of 4th degree}\nonumber\\
 \fl\label{eq:b+}
 b_+ &=&\frac{N_{21}-N_{12}}{2N_{11}}=\frac{p_2(\zeta)}{p_4(\zeta)},
 \hspace{0.895cm} p_2:\ \mbox{real polynomial of 2nd degree}\\
 \fl\label{eq:E+2}
 |\E_+|^2 &=& \frac{N_{22}}{N_{11}}=\frac{q_4(\zeta)}{p_4(\zeta)},
 \hspace{2.08cm} q_4:\ \mbox{real monic polynomial of 4th degree}\\
 \fl\label{eq:Phi+}
 \Phi_+ &=& \frac{N_{13}}{N_{11}}=\frac{p_3(\zeta)}{p_4(\zeta)},
 \hspace{2.06cm} p_3:\ \mbox{complex polynomial of 3rd degree}\\
 \fl\label{eq:Phi+2}
 |\Phi_+|^2 &=& \frac{N_{12}+N_{21}}{2N_{11}}=\frac{q_2(\zeta)}{p_4(\zeta)},\
 \hspace{0.79cm} q_2:\ \mbox{real polynomial of 2nd degree}\\
 \fl\label{eq:Phi+E+}
 \Phi_+\bar\E_+ &=& -\frac{N_{23}}{N_{11}}=\frac{q_3(\zeta)}{p_4(\zeta)}.
 \hspace{1.64cm} q_3:\ \mbox{complex polynomial of 3rd degree}
\end{eqnarray}
All polynomials in the above formulae can be given explicitly, but the coefficients are rather lengthy expressions depending on the parameters $\E_A,\dots,\E_D$, $\Phi_A,\dots,\Phi_D$, $K_A,\dots,K_D$, $\Omega_1,\ \Omega_2$. Also note that the two polynomials $p_2$ and $q_2$ initially appear to be of third degree, but in both cases the leading coefficients are proportional to the left-hand side of \eqref{eq:parcond} and hence vanish as a consequence of the parameter conditions.

Firstly, we construct $\E_+$ using \eqref{eq:f+}, \eqref{eq:b+}, \eqref{eq:Phi+2}, 
\begin{equation}\label{eq:E+formula}
  \E_+=f_+-|\Phi_+|^2+\ii b_+=\frac{\pi_4-q_2+\ii p_2}{p_4},
\end{equation}
and compare $|\E_+|^2$ as obtained from this expression with \eqref{eq:E+2}. This leads to the condition
\begin{equation}
 (\pi_4-q_2+\ii p_2)(\pi_4-q_2-\ii p_2)=q_4p_4.
\end{equation}
Comparing zeros of both sides and using that the two terms on the left-hand side are complex conjugate polynomials, and the factors on the right-hand side are real polynomials, we observe that each bracket on the left-hand side has two zeros of $q_4$ and two zeros of $p_4$. Hence, in the expression \eqref{eq:E+formula} for $\E_+$, two linear factors cancel and we actually have
\begin{equation}\label{eq:E+}
 \E_+=\frac{\pi_2(\zeta)}{r_2(\zeta)},
\end{equation}
where $\pi_2$ and $r_2$ are complex monic polynomials of 2nd degree. 

Secondly, we use the expression \eqref{eq:Phi+} for $\Phi_+$ to calculate $|\Phi_+|^2$ and compare the result with \eqref{eq:Phi+2}. In this way, we obtain the condition
\begin{equation}
 p_3\bar p_3=q_2p_4.
\end{equation}
Similarly to the above discussion, we conclude that $p_3$ has two zeros of $p_4$ and one of $q_2$. Hence \eqref{eq:Phi+} simplifies to
\begin{equation}\label{eq:Phi+1}
 \Phi_+=\frac{\pi_1(\zeta)}{R_2(\zeta)},
\end{equation}
where $\pi_1$ and $R_2$ are complex polynomials of first and second degrees, respectively, which we choose such that $R_2$ is a monic polynomial.

Finally, we show that, in fact, $R_2=r_2$, i.e.\ $\Phi_+$ has the same denominator as $\E_+$. For that purpose, we use \eqref{eq:E+} and \eqref{eq:Phi+1} to construct $\Phi_+\bar\E_+=\frac{\pi_1\bar\pi_2}{R_2 \bar r_2}$ and compare with \eqref{eq:Phi+E+}, which shows that
\begin{equation}\label{eq:p4a}
 p_4=R_2\bar r_2.
\end{equation}
(Since both sides in the previous equation are monic polynomials, we indeed have equality and not just proportionality.)
We also use \eqref{eq:Phi+1} to obtain $|\Phi_+|^2=\frac{\pi_1\bar\pi_1}{R_2\bar R_2}$, which, together with \eqref{eq:Phi+2}, implies that 
\begin{equation}\label{eq:p4b}
 p_4=R_2\bar R_2
\end{equation}
is another representation of $p_4$. Combining \eqref{eq:p4a} and \eqref{eq:p4b}, we immediately confirm that $R_2=r_2$ must hold. 

Hence the previous formulae for the axis potentials finally simplify to
\begin{equation}
 \E_+=\frac{\pi_2(\zeta)}{r_2(\zeta)},\quad
 \Phi_+=\frac{\pi_1(\zeta)}{r_2(\zeta)},
\end{equation}
i.e.\ the axis values are given in terms of complex polynomials $\pi_2$, $r_2$ and $\pi_1$ of the indicated degrees, where $\pi_2$ and $r_2$ are monic polynomials.

\section{Discussion\label{sec:discussion}}

We have derived the most general axis data for candidate solutions that could describe axisymmetric and stationary two-black-hole configurations with non-extremal rotating and charged black holes. Necessarily, the axis values of the Ernst potentials must be of the form
\begin{equation}
 \E_+(\zeta)=\frac{(\zeta-c_1)(\zeta-c_2)}{(\zeta-d_1)(\zeta-d_2)},\quad
 \Phi_+(\zeta)=\frac{e_1\zeta+e_2}{(\zeta-d_1)(\zeta-d_2)}
\end{equation}
with complex constants $c_i$, $d_i$, $e_i$, $i=1,2$, which corresponds to 12 real degrees of freedom. Note that we can immediately reduce the available degrees of freedom by comparing the asymptotic expansions of these data,
\begin{equation}
 \E_+(\zeta)=1-\frac{c_1+c_2-d_1-d_2}{\zeta}+\mathcal O(\zeta^{-2}),\quad
 \Phi_+(\zeta)= \frac{e_1}{\zeta}+\mathcal O(\zeta^{-2}),
\end{equation}
with the general behaviour of the axis potentials in an asymptotically flat spacetime with total (ADM) mass $M$ and charge $Q$, and without NUT parameter and without magnetic charge,
\begin{equation} 
 \E_+(\zeta)=1-\frac{2M}{\zeta}+\mathcal O(\zeta^{-2}),\quad
 \Phi_+(\zeta)=\frac{Q}{\zeta}+\mathcal O(\zeta^{-2}).
\end{equation}
Obviously, we require the constraints
\begin{equation}
 \Im(c_1+c_2-d_1-d_2)=0,\quad \Im(e_1)=0,
\end{equation}
which leaves us with 10 degrees of freedom.

Note that the boundary data \eqref{eq:bdata1} and \eqref{eq:bdata2a}, \eqref{eq:bdata2b} for the existing 8-parametric exact candidate solutions discussed in Sec.~\ref{sec:intro} are all of the above form. Hence it remains to decide in future investigations whether these solutions do indeed describe physically acceptable equilibrium configurations, i.e.\ solutions for which the regularity requirements (i)--(v) from Sec.~\ref{sec:intro} are all satisfied for particular choices of the parameters. Moreover, it should be studied whether slightly larger solution classes (with the above-mentioned 10 degrees of freedom) need to be considered as well.

\section*{Acknowledgments}
I would like to thank Dominic Searles for commenting on the manuscript. 
 


\end{document}